\documentclass[twocolumn]{aastex701}

\newboolean{showrevisions}
\setboolean{showrevisions}{true} % 切换 true/false
\newcommand{\revise}[1]{%
  \ifthenelse{\boolean{showrevisions}}{\textcolor{red}{#1}}{#1}%
}

\shorttitle{Dense CSM Interaction in SNe Ibn, SNe Icn, and FBOTs}
\shortauthors{Ni et al.}
\graphicspath{{./}{figures/}}
\usepackage{amsmath}
\usepackage{threeparttable}
\usepackage{booktabs}
\usepackage{changepage} 
\usepackage{adjustbox}

\graphicspath{{./}{figures/}}
\begin{document}
\title{Mapping the Dense Circumstellar Environments of SNe Ibn, SNe Icn, and Fast Blue Optical Transients}
% \title{Light-Curve Connections among SNe Ibn, SNe Icn, and Fast Blue Optical Transients from Unified CSM-Interaction Modeling}

\author[orcid=0009-0000-0015-9124]{Kang-Rui Ni}
\affiliation{Institute of Astrophysics, Central China Normal University, Wuhan 430079, China; \url{liuld@ccnu.edu.cn;yuyw@ccnu.edu.cn}}
\affiliation{Laboratory for Compact Object Astrophysics and Astronomical Technology, Central China Normal University, Wuhan 430079, China}
\affiliation{Education Research and Application Center, National Astronomical Data Center, Wuhan 430079, China}
\email{nikangrui@mails.ccnu.edu.cn}

\author[orcid=0009-0000-2423-6825]{Yu-Hao Zhang}
\affiliation{Institute of Astrophysics, Central China Normal University, Wuhan 430079, China; \url{liuld@ccnu.edu.cn;yuyw@ccnu.edu.cn}}
\affiliation{Laboratory for Compact Object Astrophysics and Astronomical Technology, Central China Normal University, Wuhan 430079, China}
\affiliation{Education Research and Application Center, National Astronomical Data Center, Wuhan 430079, China}
\email{zhang-yh@mails.ccnu.edu.cn}

\author[0000-0002-8708-0597]{Liang-Duan Liu}
\affiliation{Institute of Astrophysics, Central China Normal University, Wuhan 430079, China; \url{liuld@ccnu.edu.cn;yuyw@ccnu.edu.cn}}
\affiliation{Laboratory for Compact Object Astrophysics and Astronomical Technology, Central China Normal University, Wuhan 430079, China}
\affiliation{Education Research and Application Center, National Astronomical Data Center, Wuhan 430079, China}
\email{liuld@ccnu.edu.cn}

\author[0000-0002-1067-1911]{Yun-Wei Yu}
\affiliation{Institute of Astrophysics, Central China Normal University, Wuhan 430079, China; \url{liuld@ccnu.edu.cn;yuyw@ccnu.edu.cn}}
\affiliation{Laboratory for Compact Object Astrophysics and Astronomical Technology, Central 
China Normal University, Wuhan 430079, China}
\affiliation{Education Research and Application Center, National Astronomical Data Center, Wuhan 430079, China}
\email{yuyw@ccnu.edu.cn}

\author[0000-0002-9092-0593]{Ji-an Jiang}
\affiliation{Department of Astronomy, University of Science and Technology of China, Hefei 230026, China}
\affiliation{National Astronomical Observatory of Japan, 2-21-1 Osawa, Mitaka, Tokyo 181-8588, Japan}
\email{jian.jiang@ustc.edu.cn}

\begin{abstract}
SNe Ibn and SNe Icn are stripped-envelope explosions whose optical emission is commonly linked to interaction with H-poor circumstellar material (CSM), whereas fast blue optical transients (FBOTs) form an observational class of rapidly evolving, blue, and luminous events with diverse proposed power sources. We present a uniform comparison of these transients to test whether they are separated in optical light-curve and fitted physical parameter space. We compile multiband optical light curves of 25 SNe Ibn, SNe Icn, and FBOTs, measure same-band observables with Gaussian-process reconstructions, and model the data with the unified \texttt{TransFit-CSM} framework. In the observed (g)-band peak-luminosity--rise-time and decline--rise-time planes, the three classes are not cleanly separated: FBOTs preferentially occupy the luminous and rapidly evolving end of the distribution, but show limited overlap with part of the Ibn/Icn population. Their extinction-corrected peak colors span a broadly overlapping blue region, with FBOTs extending to bluer colors. Unified CSM-interaction fits, including shock heating and an effective inner heating component, yield overlapping CSM and ejecta parameter distributions. These results indicate that the optical light curves of SNe Ibn, SNe Icn, and at least some FBOTs can be compared within a common dense-CSM interaction framework, while the most extreme FBOTs may still require additional power sources or non-thermal components.
\end{abstract}


\section{Introduction}

Large-scale time-domain surveys have revealed a growing population of rapidly evolving optical transients with short durations, blue continua, and high peak luminosities \citep{Drout2014,Pursiainen2018}. These discoveries have broadened the observed diversity of stellar explosions and have also blurred the empirical boundaries among traditional transient classes \citep{Inserra2019}. In particular, several studies have noted photometric and spectroscopic similarities between some fast blue optical transients (FBOTs) and SNe Ibn, suggesting that dense circumstellar material (CSM) interaction may contribute to the optical emission of at least a subset of FBOT-like events \citep{Pellegrino2022a,Xiang2021}.

SNe Ibn and SNe Icn are stripped-envelope supernovae most directly associated with interaction in dense H-poor circumstellar environments. SNe Ibn are characterized by rapidly evolving light curves and narrow He emission lines superposed on hot blue continua, indicating interaction with dense He-rich, H-poor CSM. They are commonly interpreted as interaction-powered explosions preceded by intense mass loss shortly before core collapse \citep{Hosseinzadeh2017}. SNe Icn, by contrast, are identified primarily by narrow C/O features, pointing to interaction with H-poor, He-poor, and C/O-rich CSM \citep{GalYam2022,Davis2023,Nagao2023}. Although the progenitor channels of both classes remain uncertain, these events demonstrate that rapidly evolving stripped-envelope explosions can arise from interaction with chemically unusual dense CSM.

FBOTs are less clearly defined in physical terms. Observationally, they are selected as rapidly evolving, blue, and luminous transients, and some well-studied events show strong X-ray or radio emission \citep{Perley2019,Margutti2019}. An even more extreme subset has been referred to as ``fast blue ultraluminous transients'' (FBUTs), characterized by ultrahigh UV/blue-optical luminosities and short rise times comparable to or shorter than those of ordinary FBOTs \citep{Jiang2022}. Their proposed power sources include dense CSM interaction, central engines, explosions with low ejecta masses, and other scenarios \citep[e.g.,][]{Drout2014,Yu2015,Perley2019,Margutti2019,Liu2022}. FBOTs should therefore be regarded as an observationally defined ensemble rather than as a class with a single established physical origin. A useful question is not whether all FBOTs belong to one existing category, but whether some FBOTs overlap with SNe Ibn/Icn in observational and physical parameter space.

A direct comparison among SNe Ibn, SNe Icn, and FBOTs is challenging because previous studies have often relied on heterogeneous samples, class-specific analyses, and different light-curve models or parameter definitions 
\citep{Drout2014,Pursiainen2018,Maeda2022,Pellegrino2022a,Pellegrino2022b,Ho2023}. 
Many works focus on individual events or on one class at a time, so the physical parameters reported in the literature are not always placed in a common coordinate system 
\citep{Xiang2021,GalYam2022,Perley2022,Nagao2023}. 
As a result, it remains difficult to determine whether the apparent similarities among these transients reflect a genuine physical connection or only a phenomenological overlap in the fast, blue, and luminous region of transient parameter space. 
A uniform sample-level analysis is therefore needed, using the same empirical measurements, the same fitting procedure, and the same model parameterization.

The \texttt{TransFit}\footnote{\url{https://github.com/CCNUastro/TransFit}.} framework models supernova and optical-transient light curves by solving the time-dependent radiative-diffusion equation \citep{Liu2025TransFit}. Its CSM-interaction extension, \texttt{TransFit-CSM}, couples thin-shell ejecta--CSM interaction dynamics to radiative diffusion through an optically thick CSM shell \citep{Zhang2026TransFitCSM}. In this formulation, the shock-heating region and the effective photon-escape path evolve self-consistently as the interaction proceeds. The model therefore provides a common physical parameterization for comparing interaction-powered transients with different CSM masses, radial extents, shock-deposition histories, and diffusion timescales.

In this work, we examine the connections among SNe Ibn, SNe Icn, and FBOTs using both empirical light-curve measurements and unified CSM-interaction modeling. We first use band-by-band Gaussian-process reconstructions to measure peak luminosities, rise timescales, decline proxies, and peak colors on a common observational basis. We then adopt \texttt{TransFit-CSM} as the baseline framework and use a unified shock+$^{56}$Ni source-injection model to fit the multiband light curves of all retained events. This approach allows us to compare the three classes in a shared fitted parameter space. Our goal is not to assume a common origin for all events, but to test whether their optical light curves and fitted CSM/ejecta parameters show continuous or overlapping behavior under a uniform interaction-based framework.

The structure of this paper is as follows. Section~\ref{sec:data_obs} describes the sample selection, Gaussian-process reconstructions, and empirical light-curve measurements. Section~\ref{sec:model} presents the unified CSM-interaction light-curve model. Section~\ref{sec:results} gives the representative multiband fits and compares the fitted parameter distributions. Section~\ref{sec:discussion} discusses the implications, limitations, and conclusions.

\section{Observational Sample and Empirical Light-curve Measurements}
\label{sec:data_obs}

\begin{table}[t]
\centering
\caption{Observational Properties of the Adopted Sample}
\label{tab:obsinfo_final}
\scriptsize
\renewcommand{\arraystretch}{1.08}
\setlength{\tabcolsep}{3.5pt}

\begin{threeparttable}
\begin{tabular*}{\columnwidth}{@{\extracolsep{\fill}}l c c c@{}}
\toprule
Source & $z$ &
\begin{tabular}{@{}c@{}}$E(B-V)_{\rm MW}$\\(mag)\end{tabular}
& Ref. \\
\midrule
\multicolumn{4}{l}{\textbf{SNe Ibn}} \\
SN2010al  & 0.0172  & 0.040 & (1) \\
SN2014av  & 0.03015 & 0.015 & (2) \\
SN2018jmt & 0.032   & 0.105 & (3) \\
SN2019cj  & 0.0444  & 0.016 & (3) \\
SN2019myn & 0.100   & 0.043 & (4) \\
SN2019uo  & 0.02045 & 0.011 & (5) \\
SN2020bqj & 0.068   & 0.020 & (6) \\
SN2020taz & 0.0494  & 0.091 & (7) \\
SN2021jpk & 0.038   & 0.018 & (8) \\
SN2024aej & 0.063   & 0.055 & (7) \\

\addlinespace[2pt]
\multicolumn{4}{l}{\textbf{SNe Icn}} \\
SN2019hgp & 0.0641  & 0.019 & (9)  \\
SN2019jc  & 0.01948 & 0.057 & (10) \\
SN2021ckj & 0.141   & 0.047 & (11) \\
SN2021csp & 0.084   & 0.026 & (12) \\
SN2022ann & 0.04938 & 0.033 & (13) \\

\addlinespace[2pt]
\multicolumn{4}{l}{\textbf{FBOTs}} \\
AT2018cow & 0.01414 & 0.076 & (14) \\
AT2020bot & 0.197   & 0.024 & (4)  \\
AT2020mrf & 0.1353  & 0.018 & (15) \\
AT2020xnd & 0.2433  & 0.069 & (16) \\
CSS161010 & 0.034   & 0.082 & (17) \\
LSQ13ddu  & 0.05779 & 0.008 & (18) \\
PS1-10bjp & 0.113   & 0.049 & (19) \\
PS1-11qr  & 0.324   & 0.017 & (19) \\
PS1-12bv  & 0.405   & 0.010 & (19) \\
SN2018gep & 0.03154 & 0.009 & (20) \\
\bottomrule
\end{tabular*}

\begin{tablenotes}[flushleft]
\scriptsize
\item \textit{Note.} Source classifications are indicated by the boldface group headings. The listed $E(B-V)_{\rm MW}$ values are those adopted in the light-curve compilation used for the present analysis.
\item \textit{References.}
(1) \citealt{Pastorello2015a};
(2) \citealt{Pastorello2016};
(3) \citealt{Wang2024};
(4) \citealt{Ho2023};
(5) \citealt{Gangopadhyay2020};
(6) \citealt{Kool2021};
(7) \citealt{Wang2025};
(8) \citealt{Pellegrino2022a};
(9) \citealt{GalYam2022};
(10) \citealt{Pellegrino2022b};
(11) \citealt{Nagao2023};
(12) \citealt{Perley2022};
(13) \citealt{Davis2023};
(14) \citealt{Perley2019};
(15) \citealt{Yao2022};
(16) \citealt{Perley2021};
(17) \citealt{Gutierrez2024};
(18) \citealt{Clark2020};
(19) \citealt{Drout2014};
(20) \citealt{Ho2019}.
\end{tablenotes}
\end{threeparttable}
\end{table}

In this section, we describe the observational data set used for the cross-class comparison of SNe Ibn, SNe Icn, and FBOTs. The goal is to place the three classes on a common empirical basis before applying the unified \texttt{TransFit-CSM} modeling. We first define the retained sample and then describe the Gaussian-process light-curve reconstruction used to measure the peak luminosities, rise timescales, decline proxies, and peak colors.

\subsection{Sample Definition}
\label{subsec:Sample}

We compiled multiband light curves of SNe Ibn, SNe Icn, and FBOTs from the literature to construct a common observational basis for the cross-class comparison. The parent sample was defined using the following criteria:
\begin{enumerate}
\item the event has a published classification as an SN Ibn, SN Icn, or FBOT, or an equivalent classification widely adopted in public transient databases;
\item optical photometry is available in multiple bands and covers the light-curve peak sufficiently to constrain the overall evolution;
\item at least one pre-maximum measurement or constraining upper limit is available to help determine the rise behavior or explosion-time offset; 
\item the redshift or distance, together with the Milky Way foreground extinction, is available for luminosity and color calibration.
\end{enumerate}

These criteria retain events whose optical light curves can be placed on a common observational scale and subsequently modeled within the unified \texttt{TransFit-CSM} framework. We do not require identical cadence, filter coverage, or same-band pre-peak sampling for all objects, because such requirements would substantially reduce the available sample. Instead, additional quality cuts are applied below when constructing the more restrictive empirical subsamples used for same-band comparisons. In particular, the GP-based measurements of $\tau_{\rm rise,g}$ and $\Delta m_{10,g}$ are restricted to events whose observed $g$-band reconstructions provide well-constrained peak and pre-peak behavior. The retained parent sample, including source names, redshifts, and the Milky Way foreground extinction $E(B-V)_{\rm MW}$, is summarized in Table~\ref{tab:obsinfo_final}.

\subsection{Gaussian-process Reconstruction and Same-band Observables}

To compare SNe Ibn, SNe Icn, and FBOTs on a weakly model-dependent observational basis, we measured a set of empirical light-curve quantities from Gaussian-process (GP) reconstructions. These quantities include the peak epoch $t_{\rm peak}$, the peak monochromatic luminosity $\nu L_{\nu,\rm peak}$, the rise timescale $\tau_{\rm rise}$, and the fixed-epoch decline proxy $\Delta m_{10}$.

For each band, $t_{\rm peak}$ and $\nu L_{\nu,\rm peak}$ were determined from the GP-reconstructed light curve. We define $\tau_{\rm rise}$ as the time interval between $t_{\rm peak}$ and the pre-peak crossing of $F_{\rm peak}/e$, following the common use of a $1/e$ threshold to characterize supernova light-curve timescales \citep[e.g.,][]{Nicholl2015}. The post-peak decline proxy is defined as
\begin{equation}
\Delta m_{10} = m(t_{\rm peak}+10~ {\rm day}) - m_{\rm peak},
\end{equation}
where $\Delta m_{10}$ is used only as a fixed-epoch same-band decline measure, rather than as a physical decline timescale.

The GP reconstructions were performed independently for each optical band using the \texttt{george} package \citep{Ambikasaran2016}. For the cross-object empirical comparison, we adopted the observed $g$ band as a common reference band. This choice reduces filter-dependent differences in the measured light-curve quantities while retaining a sufficiently large subsample. Accordingly, $\nu L_{\nu,\rm peak,g}$, $\tau_{\rm rise,g}$, and $\Delta m_{10,g}$ were measured from the reconstructed $g$-band light curves when the corresponding quantities were constrained by the GP posterior.

\begin{figure}[t]
    \centering
    \includegraphics[width=0.95\linewidth]{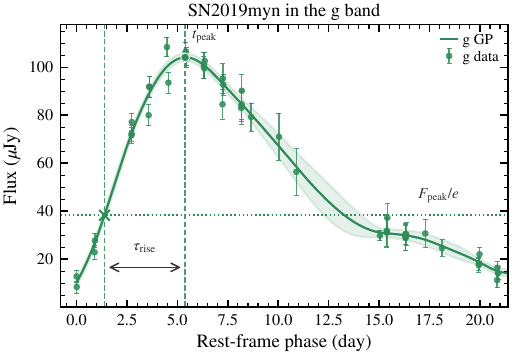}
\caption{Gaussian-process reconstruction of the $g$-band light curve of SN~2019myn. The points with error bars show the observed $g$-band photometry, while the solid curve with shaded region represents the GP median and 1$\sigma$ confidence interval. The vertical dashed lines indicate the peak epoch $t_{\rm peak}$ and the pre-peak crossing of $F_{\rm peak}/e$, and the horizontal dotted line marks $F_{\rm peak}/e$. The rise timescale $\tau_{\rm rise}$ is defined as the time interval between the pre-peak $F_{\rm peak}/e$ crossing and the peak.}
    \label{fig:sn2019myn_g_definition}
\end{figure}

In Figure~\ref{fig:sn2019myn_g_definition}, we show the GP reconstruction of the $g$-band light curve for SN~2019myn. Observed photometric points with error bars are overlaid with the GP median and 1$\sigma$ confidence interval. The peak epoch, $t_{\rm peak}$, and the pre-peak flux level $F_{\rm peak}/e$ are marked with vertical dashed lines, and the rise timescale, $\tau_{\rm rise}$, is defined as the time interval between these two epochs. This illustration clarifies the measurement of empirical light-curve observables used for same-band comparisons of SNe Ibn, SNe Icn, and FBOTs.

\subsection{Empirical Light-curve Properties from GP Reconstructions}
\label{sec:gp_obs_continuity}

\begin{figure*}[t]
\centering
\includegraphics[width=0.96\textwidth]{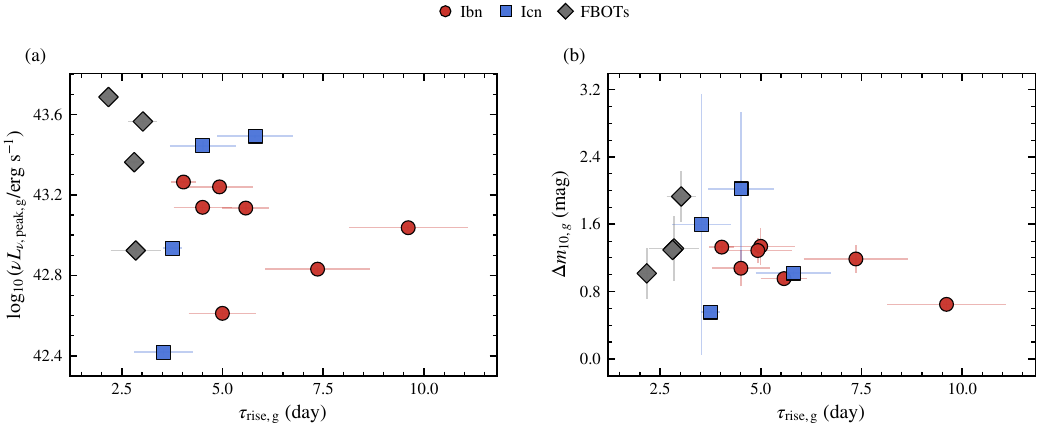}
\caption{Same-band $g$-band GP-derived empirical observables for the 15-object subsample. Panel (a) shows the peak luminosity $\log_{10}(\nu L_{\nu,\rm peak,g})$ versus the rise timescale $\tau_{\rm rise,g}$. Panel (b) shows the post-peak 10-day decline proxy $\Delta m_{10,g}$ as a function of $\tau_{\rm rise,g}$. SNe Ibn (red circles), SNe Icn (blue squares), and FBOTs (gray diamonds) are plotted with their associated uncertainties. The overlap coefficients (OVLs), estimated using kernel density estimation (KDE), between FBOTs and the combined Ibn/Icn distribution are ${\rm OVL}=0.10^{+0.11}_{-0.07}$ in panel~(a) and ${\rm OVL}=0.10^{+0.10}_{-0.06}$ in panel~(b).}
\label{fig:gp_observables_continuity_2panel}
\end{figure*}

We examine the GP–derived light-curve observables of SNe Ibn, SNe Icn, and FBOTs in the common observed $g$ band. Figure~\ref{fig:gp_observables_continuity_2panel} compares the same-band empirical subsample in two planes: $\log_{10}(\nu L_{\nu,\rm peak,g})$ versus $\tau_{\rm rise,g}$ and $\Delta m_{10,g}$ versus $\tau_{\rm rise,g}$. The 25 objects listed in Table~\ref{tab:obsinfo_final} constitute the full retained sample, whereas Figure~\ref{fig:gp_observables_continuity_2panel} uses a more restrictive same-band subsample. To enter this subsample, an object must have finite GP-derived measurements of $t_{\rm peak,g}$, $\nu L_{\nu,\rm peak,g}$, and $\tau_{\rm rise,g}$, with the pre-peak $F_{\rm peak}/e$ crossing constrained by the GP posterior. Fifteen objects satisfy these requirements. The KDE-based overlap coefficients reported throughout this work are calculated as described in Appendix~\ref{app:kde-ovl}.

Figure~\ref{fig:gp_observables_continuity_2panel} shows that the three classes are not cleanly separated in the observed $g$-band peak-luminosity--rise-time plane. FBOTs preferentially occupy the high-luminosity and short-rise region, whereas SNe Ibn and SNe Icn extend toward longer rise times and lower peak luminosities, with partial overlap between the two populations. This behavior is qualitatively consistent with the expectation that rapidly evolving luminous transients require a short effective photon escape time and efficient conversion of kinetic or engine energy into radiation. In an interaction-powered interpretation, such conditions can be realized if the emitting region is compact, the optically thick CSM has a relatively small diffusion time, or the shock power is deposited close to the photospheric region. The location of FBOTs in this plane therefore suggests that, if their optical peaks are powered at least partly by CSM interaction, they likely correspond to the hotter and more rapidly radiating end of the interaction parameter space.

The $\Delta m_{10,g}$--$\tau_{\rm rise,g}$ plane provides a complementary view of the post-peak evolution. Although $\Delta m_{10,g}$ is not a physical decline timescale, it traces how rapidly the optical emission fades over a fixed rest-frame interval after maximum light. Objects with short $\tau_{\rm rise,g}$ and large $\Delta m_{10,g}$ are characterized by both rapid energy release before peak and rapid cooling or diffusion after peak. This combination is expected when the radiating material has a small effective diffusion mass or when the shock-heated region loses its stored thermal energy on a short timescale. Conversely, events with longer rise times and smaller $\Delta m_{10,g}$ may correspond to larger diffusion masses, more extended optically thick CSM, or a more sustained heating contribution. The broad spread in $\Delta m_{10,g}$ at a given $\tau_{\rm rise,g}$ indicates that the post-peak decline is not controlled by a single parameter, but likely reflects a combination of CSM mass, radial extent, density profile, opacity, and heating efficiency.

The partial overlap between SNe Ibn/Icn and FBOTs in both panels is therefore physically informative. It indicates that the empirical diversity of these events can be arranged along a continuous range of optical light-curve timescales and luminosities, rather than requiring three completely distinct phenomenological groups. At the same time, the preferential location of FBOTs toward shorter $\tau_{\rm rise,g}$, higher $\nu L_{\nu,\rm peak,g}$, and in some cases larger $\Delta m_{10,g}$ suggests that they represent a more extreme region of the same observational space. This motivates the use of a unified CSM-interaction framework to test whether the apparent continuity in empirical observables is also reflected in the fitted physical parameters, while allowing for the possibility that the most luminous or rapidly fading FBOTs require additional energy input or non-thermal components beyond the optical CSM-powered emission.

\subsection{Peak Colors and Optical Temperature Indicators}
\label{sec:color_temperature}

Peak optical color provides an additional observational dimension to examine whether the empirical differences among SNe Ibn, SNe Icn, and FBOTs are reflected in their maximum-light colors. Following \citet{Jin2026}, we define the peak color at the $r$-band maximum as $(g-r)_{r,\max}=g(t_{r,\max})-r(t_{r,\max})$, where $t_{r,\max}$ is obtained from the $r$-band GP reconstruction, and $g(t_{r,\max})$ and $r(t_{r,\max})$ are evaluated from the corresponding GP-interpolated light curves. This definition allows a consistent comparison of peak colors across the sample using extinction-corrected photometry. From the 25 retained events, we select objects with reliable standard $g$- and $r$-band coverage and an $r$-band maximum directly constrained by the GP reconstruction. The resulting color subsample includes 6 SNe Ibn, 4 SNe Icn, and 6 FBOTs.

\begin{figure}[t]
\centering
\includegraphics[width=0.95\linewidth]{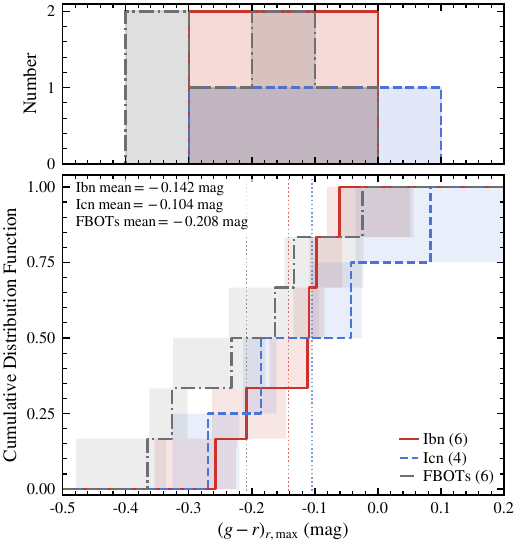}
\caption{Cumulative distributions of the extinction-corrected peak color $(g-r)_{r,\max}$ for the color subsample of SNe Ibn, SNe Icn, and FBOTs. The peak color is measured at the epoch of $r$-band maximum, $t_{r,\max}$, using GP-interpolated $g$- and $r$-band light curves. The red solid, blue dashed, and gray dash-dotted curves denote SNe Ibn, SNe Icn, and FBOTs, respectively, with shaded regions indicating the corresponding color ranges. Vertical dotted lines mark the mean color of each class. The three classes occupy a broadly overlapping blue-color region, while FBOTs extend toward bluer colors. The corresponding KDE-based OVL between FBOTs and the combined Ibn/Icn distribution is ${\rm OVL}=0.69^{+0.12}_{-0.17}$.}
\label{fig:gr_rmax_distribution}
\end{figure}

Figure~\ref{fig:gr_rmax_distribution} compares the extinction-corrected peak colors of the color subsample. All three classes occupy a blue range of $(g-r)_{r,\max}$, and their cumulative distributions show substantial overlap. The mean colors are approximately $(g-r)_{r,\max}\simeq -0.14$, $-0.10$, and $-0.21$ mag for SNe Ibn, SNe Icn, and FBOTs, respectively. Thus, FBOTs are shifted toward the bluer side of the distribution, although the small subsample size, especially for SNe Icn, prevents a statistically significant separation from being established. This color behavior is consistent with the same-band light-curve comparison: the three classes are not cleanly segregated, but FBOTs tend to occupy the hotter and more rapidly evolving end of the observed distribution.

\section{Unified CSM-interaction Light-curve Model}
\label{sec:model}

We adopt \texttt{TransFit-CSM} as the baseline light-curve modeling framework. The original \texttt{TransFit-CSM} model couples thin-shell ejecta--CSM interaction dynamics with time-dependent radiative diffusion in an optically thick CSM shell \citep{Zhang2026TransFitCSM}. In this framework, both the shock-heating location and the effective photon diffusion path evolve with time, making it more suitable for interaction-powered light curves than semi-analytic models that assume centrally deposited heating and a fixed effective diffusion timescale \citep[e.g.,][]{Chatzopoulos2012,Chatzopoulos2013}.

In this work, we retain the \texttt{TransFit-CSM} interaction dynamics and modify the heating prescription. Specifically, the shock-boundary heating in the original model is recast as a localized source term that propagates with the forward shock, and an effective inner $^{56}$Ni radioactive-heating source is added. Shock heating and radioactive heating are therefore treated within the same time-dependent diffusion equation, while the shock trajectory is still determined by the original thin-shell dynamics.

The CSM is modeled as a power-law shell extending from $R_{\rm in}$ to $R_{\rm CSM}$. In the unified fits below, we adopt the fiducial wind-like case $s=2$, corresponding to $\rho_{\rm csm}\propto r^{-2}$. The ejecta are described by a homologously expanding broken power-law density profile with an inner shallow component and an outer steep envelope. Unless otherwise stated, the fixed structural parameters are the same for all objects, so that SNe Ibn, SNe Icn, and FBOTs are compared within a common parameterization.

The shock heating is computed from the forward-shock power and multiplied by a thermalization efficiency $\epsilon_{\rm sh}$. In contrast to the original boundary-injection form of \texttt{TransFit-CSM}, this heating is written here as a localized volumetric source moving with the shock. The $^{56}$Ni heating is placed in the inner part of the shell and follows the standard radioactive decay power, with gamma-ray leakage included through a deposition factor. In the present framework, this radioactive component should be interpreted as an effective inner heating source rather than as a claim that all events are primarily powered by large amounts of $^{56}$Ni.

The bolometric luminosity is obtained from the diffusive flux at the outer boundary. To compare the model with multiband data, we introduce an effective photospheric radius and temperature through the Stefan--Boltzmann relation,
\begin{equation}
L_{\rm bol}
=
4\pi\sigma_{\rm SB}R_{\rm ph}^{2}T_{\rm ph}^{4}.
\end{equation}
The photospheric emission is approximated as a blackbody and converted to observed-frame fluxes and AB magnitudes in each band using the source redshift and luminosity distance. This procedure provides both bolometric and multiband light curves from a single set of physical parameters.

\begin{figure*}[t]
    \centering
    \includegraphics[width=0.96\textwidth]{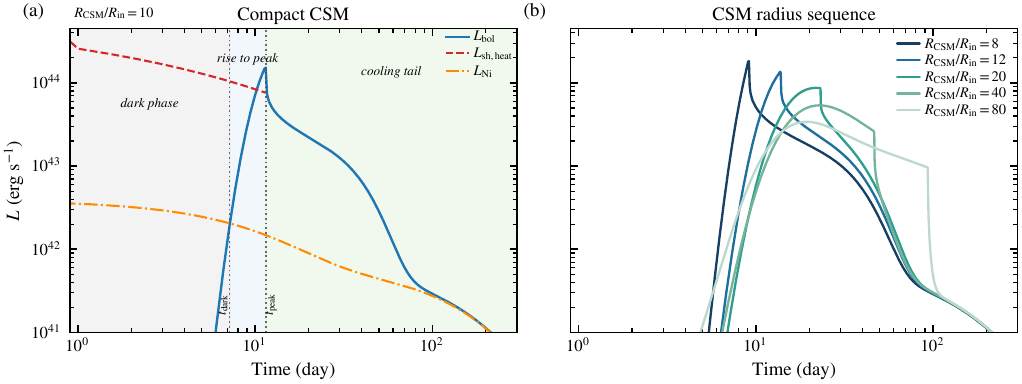}
    \caption{Bolometric light-curve morphologies produced by the unified shock+$^{56}$Ni source-injection model. Panel~(a) shows a representative compact-CSM configuration, in which the blue solid line denotes the bolometric luminosity $L_{\rm bol}$, the red dashed line denotes the thermalized shock-heating power $L_{\rm sh,heat}$, and the orange dash-dotted line denotes the radioactive-heating input $L_{\rm Ni}$. Vertical dotted lines mark $t_{\rm dark}$ and $t_{\rm peak}$. Panel~(b) shows a sequence of bolometric light curves obtained by increasing the CSM extent while keeping all other model parameters fixed. In both panels, we adopt $M_{\rm ej}=5\,M_\odot$, $E_{\rm SN}=10^{51}\,{\rm erg}$, $M_{\rm CSM}=1\,M_\odot$, $R_{\rm in}=500\,R_\odot$, $s=2$, $\kappa=0.2\,{\rm cm^2\,g^{-1}}$, $\epsilon_{\rm sh}=0.8$, $M_{\rm Ni}=0.05\,M_\odot$.}
    \label{fig:model_bol_sequence}
\end{figure*}

Figure~\ref{fig:model_bol_sequence} illustrates the physical origin of different bolometric light-curve morphologies in the unified shock+$^{56}$Ni model. In panel~(a), the compact CSM configuration ($R_{\rm CSM}/R_{\rm in}=10$) shows an initial “dark phase” during which photons are trapped within the optically thick CSM and the emergent luminosity lags behind the instantaneous shock and radioactive heating. The rise to peak is governed by the diffusion timescale through the CSM, while the peak luminosity is set primarily by the shock power at the outer layers. The post-peak decline reflects a combination of decreasing shock energy deposition, adiabatic cooling of the ejecta, and the residual contribution from radioactive decay. Panel~(b) highlights the effect of increasing the CSM radial extent on the light curve. Increasing $R_{\rm CSM}/R_{\rm in}$ changes both the radius at which shock energy is deposited and the optical-depth structure through which photons escape. 
The resulting changes in peak time, peak luminosity, and light-curve width should therefore be interpreted as the combined effect of shock propagation and time-dependent diffusion, rather than as a change in a single constant diffusion timescale. 
This point is particularly important for a fixed $M_{\rm CSM}$ and $s=2$ shell, because redistributing the same CSM mass over a larger radial extent modifies the density and optical-depth profiles.

These model examples motivate the use of a unified CSM-interaction
framework for the full sample. Variations in CSM mass, radial extent,
opacity, ejecta properties, and heating efficiency can produce a
continuous range of optical rise times, peak luminosities, and decline
rates. Consequently, the model provides a controlled way to test whether
SNe Ibn, SNe Icn, and FBOTs occupy overlapping regions of fitted
interaction parameter space. At the same time, a successful optical
light-curve fit does not prove that CSM interaction is the only power
source in a given event; rather, it shows whether the observed optical
evolution can be described within a common interaction-based thermal
framework.

\begin{table}[t]
\centering
\caption{Physical Fit Parameters and Priors Used in the Model}
\label{tab:model_priors}
\footnotesize
\renewcommand{\arraystretch}{1.15}

\begin{tabular*}{\columnwidth}{@{\extracolsep{\fill}}l c c c@{}}
\toprule
Parameter & Prior & Min & Max \\
\midrule
$M_{\rm CSM}/M_\odot$                & Log-flat & $0.01$   & $3.00$  \\
$M_{\rm ej}/M_\odot$                 & Log-flat & $0.20$   & $15.00$ \\
$E_{\rm SN}/(10^{51}\,{\rm erg})$    & Log-flat & $0.10$   & $30.00$ \\
$R_{\rm CSM}/(10^4\,R_\odot)$        & Log-flat & $0.50$   & $10.00$ \\
$\kappa/({\rm cm^2\,g^{-1}})$        & Log-flat & $0.03$   & $0.35$  \\
$\epsilon_{\rm sh}$                  & Flat     & $0.05$   & $0.90$  \\
$M_{\rm Ni}/M_\odot$                 & Log-flat & $0.0001$ & $0.50$  \\
$T_{\rm floor}/(10^3\,{\rm K})$      & Flat     & $3.00$   & $8.00$  \\
\bottomrule
\end{tabular*}
\end{table}

The physical fit parameters and prior ranges adopted in the unified fits are listed in Table~\ref{tab:model_priors}. These parameters describe the CSM and ejecta properties, the diffusion opacity, the shock-heating efficiency, the effective radioactive-heating component. The same priors are used for all objects so that the resulting fitted parameters can be compared consistently across SNe Ibn, SNe Icn, and FBOTs.

\section{Unified Light-curve Fits and Fitted Parameter Space}
\label{sec:results}

\subsection{Representative Multiband Fits}
\label{subsec:representative-multiband-fits}

\begin{figure*}[t]
    \centering
    \includegraphics[width=\textwidth]{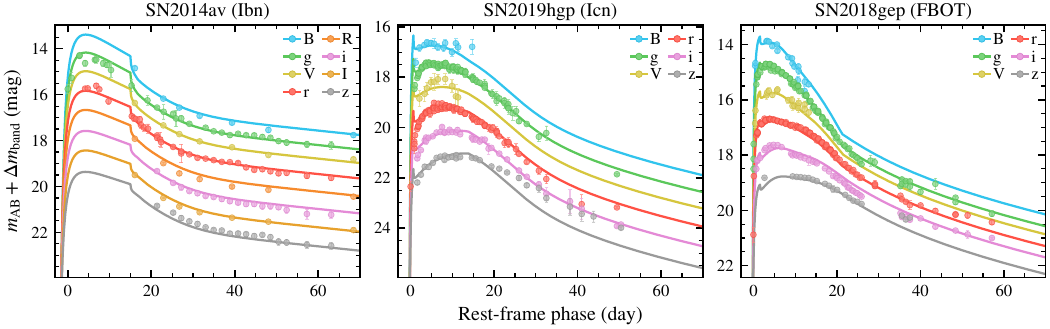}
    \caption{Representative unified \texttt{TransFit-CSM} multiband fits for one retained event from each of the Ibn, Icn, and FBOT classes: SN~2014av, SN~2019hgp, and SN~2018gep. Solid lines show the model light curves, and circles show the observed photometry.}
    \label{fig:representative-fits}
\end{figure*}

We first examine whether the optical multiband light curves of SNe Ibn, SNe Icn, and FBOTs can be described within the unified \texttt{TransFit-CSM} framework. Figure~\ref{fig:representative-fits} shows representative fits for one retained event from each class: SN~2014av, SN~2019hgp, and SN~2018gep. These examples are not intended to replace the sample-level parameter comparison below. Instead, they illustrate that the same interaction-based parameterization can reproduce the main optical light-curve morphology of events with different peak luminosities, timescales, and post-peak decline behavior.

As shown in Figure~\ref{fig:representative-fits}, the unified framework generally reproduces the main peak-region structure, the relative ordering among bands, and the dominant post-peak decline behavior in all three classes. In other words, although SNe Ibn, SNe Icn, and FBOTs span a broad range of observed light-curve morphologies, these differences do not require completely separate empirical fitting languages at the level of representative multiband modeling. At least for these illustrative cases, the unified \texttt{TransFit-CSM} framework shows substantial descriptive power across the three classes.

The corresponding object-by-object multiband fits for the full retained sample are presented in Appendix~\ref{app:full-sample-fits} (Figure~\ref{fig:full-sample-fits}). These full-sample fits extend the representative examples shown in Figure~\ref{fig:representative-fits} and provide a direct visual counterpart to the fitted parameters summarized in Table~\ref{tab:fit-parameters}.

The posterior median fit parameters for the 25 retained events are summarized in Table~\ref{tab:fit-parameters}, with all quantities reported as posterior medians and their 16th--84th percentile ranges. Table~\ref{tab:fit-parameters} provides the full set of object-by-object fit results and forms the basis for the parameter-space comparison presented in the next subsection. The main conclusion of this section is thus that the observational non-segregation identified in Section~2 does not disappear at the level of unified light-curve fitting: instead, the multiband light curves of SNe Ibn, SNe Icn, and FBOTs can be described in a comparable way within a common interaction-based thermal framework.
\begin{table*}[t]
    \centering
    \caption{Posterior Median Fit Parameters for the Adopted SNe Ibn, SNe Icn, and FBOT Sample}
    \label{tab:fit-parameters}
    \footnotesize
    \renewcommand{\arraystretch}{1.18}
    \setlength{\tabcolsep}{3.2pt}

    \begin{tabular*}{\textwidth}{@{\extracolsep{\fill}}l c c c c c c c c@{}}
        \toprule
        Source
        & $M_{\rm CSM}$
        & $M_{\rm ej}$
        & $E_{\rm SN}$
        & $R_{\rm CSM}$
        & $\kappa$
        & $\epsilon_{\rm sh}$
        & $M_{\rm Ni}$
        & $T_{\rm floor}$ \\
        & ($M_\odot$)
        & ($M_\odot$)
        & ($10^{51}\,{\rm erg}$)
        & ($10^4\,R_\odot$)
        & (${\rm cm^2\,g^{-1}}$)
        &
        & ($M_\odot$)
        & ($10^3\,{\rm K}$) \\
        \midrule
        \multicolumn{9}{l}{\textbf{SNe Ibn}} \\
        SN~2010al & $0.84^{+0.51}_{-0.23}$ & $3.98^{+2.98}_{-1.24}$ & $0.54^{+0.40}_{-0.16}$ & $1.79^{+0.05}_{-0.05}$ & $0.226^{+0.089}_{-0.093}$ & $0.264^{+0.100}_{-0.099}$ & $0.0003^{+0.0006}_{-0.0002}$ & $6.49^{+0.25}_{-0.31}$ \\
        SN~2014av & $0.32^{+0.06}_{-0.03}$ & $0.55^{+0.39}_{-0.20}$ & $0.96^{+0.31}_{-0.18}$ & $2.44^{+0.12}_{-0.13}$ & $0.298^{+0.036}_{-0.054}$ & $0.062^{+0.016}_{-0.008}$ & $0.0963^{+0.0021}_{-0.0020}$ & $6.66^{+0.07}_{-0.06}$ \\
        SN~2018jmt & $0.17^{+1.19}_{-0.01}$ & $0.21^{+0.81}_{-0.01}$ & $0.11^{+0.23}_{-0.00}$ & $1.33^{+0.03}_{-0.02}$ & $0.346^{+0.003}_{-0.273}$ & $0.534^{+0.039}_{-0.184}$ & $0.1468^{+0.0066}_{-0.0444}$ & $8.00^{+0.00}_{-0.01}$ \\
        SN~2019cj & $2.86^{+0.11}_{-0.25}$ & $0.25^{+0.02}_{-0.03}$ & $0.11^{+0.02}_{-0.00}$ & $2.07^{+0.04}_{-0.03}$ & $0.042^{+0.006}_{-0.002}$ & $0.353^{+0.018}_{-0.040}$ & $0.0015^{+0.0056}_{-0.0012}$ & $5.45^{+1.54}_{-1.59}$ \\
        SN~2019myn & $1.79^{+0.21}_{-1.03}$ & $0.28^{+0.05}_{-0.06}$ & $0.11^{+0.13}_{-0.01}$ & $1.99^{+0.32}_{-0.09}$ & $0.033^{+0.007}_{-0.002}$ & $0.461^{+0.115}_{-0.067}$ & $0.0014^{+0.0095}_{-0.0012}$ & $5.89^{+1.58}_{-1.98}$ \\
        SN~2019uo & $0.16^{+0.04}_{-0.02}$ & $4.99^{+6.15}_{-2.61}$ & $0.44^{+0.33}_{-0.19}$ & $0.92^{+0.02}_{-0.02}$ & $0.285^{+0.039}_{-0.067}$ & $0.796^{+0.074}_{-0.148}$ & $0.0270^{+0.0016}_{-0.0014}$ & $7.99^{+0.01}_{-0.03}$ \\
        SN~2020bqj & $2.56^{+0.19}_{-0.12}$ & $0.21^{+0.02}_{-0.01}$ & $3.81^{+0.16}_{-0.25}$ & $2.88^{+0.05}_{-0.06}$ & $0.263^{+0.013}_{-0.016}$ & $0.053^{+0.008}_{-0.002}$ & $0.2255^{+0.0118}_{-0.0106}$ & $8.00^{+0.00}_{-0.00}$ \\
        SN~2020taz & $1.25^{+0.45}_{-0.44}$ & $0.26^{+0.16}_{-0.05}$ & $0.65^{+0.31}_{-0.28}$ & $0.70^{+0.05}_{-0.04}$ & $0.051^{+0.025}_{-0.014}$ & $0.110^{+0.143}_{-0.043}$ & $0.0005^{+0.0016}_{-0.0003}$ & $7.03^{+0.38}_{-2.43}$ \\
        SN~2021jpk & $0.26^{+0.12}_{-0.13}$ & $0.41^{+0.30}_{-0.15}$ & $0.14^{+0.07}_{-0.03}$ & $0.87^{+0.06}_{-0.08}$ & $0.063^{+0.067}_{-0.022}$ & $0.093^{+0.082}_{-0.027}$ & $0.0495^{+0.0147}_{-0.0291}$ & $5.53^{+1.73}_{-1.79}$ \\
        SN~2024aej & $0.39^{+0.16}_{-0.22}$ & $1.26^{+3.44}_{-1.04}$ & $3.10^{+1.26}_{-1.27}$ & $1.74^{+0.11}_{-1.23}$ & $0.234^{+0.076}_{-0.088}$ & $0.069^{+0.028}_{-0.013}$ & $0.0519^{+0.3203}_{-0.0466}$ & $7.76^{+0.19}_{-0.38}$ \\

        \addlinespace[2pt]
        \multicolumn{9}{l}{\textbf{SNe Icn}} \\
        SN~2019hgp & $0.50^{+0.21}_{-0.14}$ & $0.34^{+0.15}_{-0.10}$ & $0.45^{+2.18}_{-0.14}$ & $1.46^{+0.03}_{-0.96}$ & $0.201^{+0.129}_{-0.070}$ & $0.071^{+0.035}_{-0.018}$ & $0.0243^{+0.0722}_{-0.0010}$ & $6.16^{+1.43}_{-0.24}$ \\
        SN~2019jc & $0.14^{+0.11}_{-0.07}$ & $0.90^{+5.07}_{-0.63}$ & $0.25^{+0.43}_{-0.09}$ & $0.54^{+0.06}_{-0.04}$ & $0.248^{+0.074}_{-0.131}$ & $0.098^{+0.159}_{-0.034}$ & $0.0225^{+0.0080}_{-0.0164}$ & $4.47^{+1.84}_{-0.87}$ \\
        SN~2021ckj & $0.24^{+0.08}_{-0.04}$ & $3.43^{+5.23}_{-2.48}$ & $6.63^{+7.60}_{-4.13}$ & $2.09^{+0.09}_{-0.16}$ & $0.161^{+0.028}_{-0.049}$ & $0.053^{+0.008}_{-0.002}$ & $0.2770^{+0.0109}_{-0.0087}$ & $7.96^{+0.03}_{-0.07}$ \\
        SN~2021csp & $1.76^{+0.48}_{-0.53}$ & $0.59^{+0.19}_{-0.17}$ & $0.52^{+0.26}_{-0.17}$ & $2.16^{+0.03}_{-0.04}$ & $0.043^{+0.017}_{-0.009}$ & $0.197^{+0.104}_{-0.072}$ & $0.1237^{+0.0584}_{-0.0113}$ & $7.97^{+0.02}_{-0.06}$ \\
        SN~2022ann & $0.63^{+0.38}_{-0.21}$ & $3.99^{+5.79}_{-2.69}$ & $0.41^{+0.36}_{-0.21}$ & $0.63^{+0.05}_{-0.03}$ & $0.109^{+0.043}_{-0.040}$ & $0.613^{+0.194}_{-0.231}$ & $0.0131^{+0.0044}_{-0.0037}$ & $7.96^{+0.03}_{-0.07}$ \\

        \addlinespace[2pt]
        \multicolumn{9}{l}{\textbf{FBOTs}} \\
        AT~2018cow & $0.11^{+0.00}_{-0.00}$ & $0.21^{+0.01}_{-0.00}$ & $1.21^{+0.03}_{-0.03}$ & $0.99^{+0.01}_{-0.01}$ & $0.095^{+0.003}_{-0.002}$ & $0.894^{+0.004}_{-0.010}$ & $0.3261^{+0.0037}_{-0.0038}$ & $8.00^{+0.00}_{-0.00}$ \\
        AT~2020bot & $0.27^{+0.42}_{-0.13}$ & $0.49^{+0.38}_{-0.20}$ & $0.63^{+0.59}_{-0.41}$ & $3.91^{+0.21}_{-0.19}$ & $0.123^{+0.148}_{-0.081}$ & $0.351^{+0.354}_{-0.164}$ & $0.1219^{+0.2952}_{-0.1208}$ & $5.21^{+1.59}_{-1.47}$ \\
        AT~2020mrf & $0.31^{+0.24}_{-0.13}$ & $2.31^{+5.34}_{-1.75}$ & $5.31^{+8.28}_{-3.15}$ & $0.89^{+0.51}_{-0.28}$ & $0.077^{+0.057}_{-0.033}$ & $0.269^{+0.268}_{-0.155}$ & $0.4258^{+0.0579}_{-0.0962}$ & $7.39^{+0.45}_{-0.64}$ \\
        AT~2020xnd & $0.13^{+0.01}_{-0.01}$ & $0.25^{+0.18}_{-0.04}$ & $2.09^{+0.65}_{-0.45}$ & $1.37^{+0.10}_{-0.06}$ & $0.132^{+0.027}_{-0.018}$ & $0.797^{+0.075}_{-0.143}$ & $0.4535^{+0.0295}_{-0.0397}$ & $7.99^{+0.01}_{-0.02}$ \\
        CSS161010 & $0.04^{+0.06}_{-0.02}$ & $2.10^{+5.52}_{-1.54}$ & $5.00^{+8.59}_{-3.38}$ & $1.98^{+0.14}_{-0.89}$ & $0.174^{+0.097}_{-0.071}$ & $0.590^{+0.203}_{-0.252}$ & $0.4360^{+0.0451}_{-0.1423}$ & $7.99^{+0.01}_{-0.02}$ \\
        LSQ13ddu & $0.15^{+0.06}_{-0.03}$ & $2.22^{+5.95}_{-1.70}$ & $4.34^{+7.12}_{-2.69}$ & $1.64^{+0.07}_{-0.07}$ & $0.251^{+0.067}_{-0.070}$ & $0.092^{+0.037}_{-0.027}$ & $0.2294^{+0.0156}_{-0.0143}$ & $7.96^{+0.03}_{-0.05}$ \\
        PS1-10bjp & $0.28^{+0.12}_{-0.07}$ & $1.03^{+1.20}_{-0.43}$ & $0.67^{+0.42}_{-0.21}$ & $1.05^{+0.04}_{-0.04}$ & $0.229^{+0.083}_{-0.074}$ & $0.093^{+0.039}_{-0.028}$ & $0.0546^{+0.0064}_{-0.0064}$ & $3.82^{+0.69}_{-0.51}$ \\
        PS1-11qr & $1.48^{+0.28}_{-0.31}$ & $0.25^{+0.10}_{-0.04}$ & $0.57^{+0.62}_{-0.27}$ & $1.93^{+0.14}_{-0.14}$ & $0.037^{+0.011}_{-0.005}$ & $0.247^{+0.234}_{-0.126}$ & $0.0029^{+0.0358}_{-0.0026}$ & $5.38^{+1.70}_{-1.63}$ \\
        PS1-12bv & $0.49^{+0.45}_{-0.22}$ & $0.43^{+0.67}_{-0.18}$ & $3.04^{+5.68}_{-1.89}$ & $0.75^{+1.31}_{-0.21}$ & $0.119^{+0.126}_{-0.064}$ & $0.129^{+0.213}_{-0.060}$ & $0.3426^{+0.0966}_{-0.3202}$ & $6.01^{+1.26}_{-2.01}$ \\
        SN~2018gep & $0.12^{+0.08}_{-0.04}$ & $2.43^{+2.95}_{-1.09}$ & $7.99^{+7.56}_{-3.36}$ & $0.73^{+0.01}_{-0.01}$ & $0.227^{+0.082}_{-0.093}$ & $0.212^{+0.080}_{-0.086}$ & $0.4428^{+0.0072}_{-0.0072}$ & $4.54^{+0.02}_{-0.02}$ \\
        \bottomrule
    \end{tabular*}

    \vspace{2pt}
    \begin{minipage}{0.98\textwidth}
        \footnotesize
        \textit{Note.} Reported values are posterior medians with 16th--84th percentile uncertainties from the unified CSM+$^{56}$Ni model fits. For parameters sampled in logarithmic space, the listed central values and asymmetric uncertainties are obtained after transforming the corresponding posterior percentiles into linear physical units.
    \end{minipage}
\end{table*}

\subsection{Distribution in the Fitted Parameter Space}
\label{sec:param_space_continuity}

\begin{figure*}[t]
\centering
\includegraphics[width=0.96\textwidth]{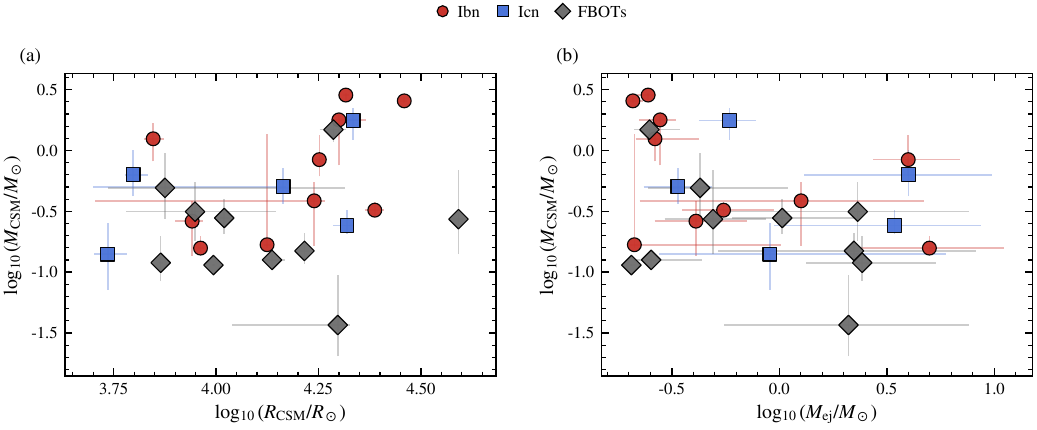}
\caption{Posterior median fitted parameters of the 25 retained events in two direct model-parameter planes. Panel~(a) shows $\log_{10}(R_{\rm CSM}/R_\odot)$ versus $\log_{10}(M_{\rm CSM}/M_\odot)$, and panel~(b) shows $\log_{10}(M_{\rm ej}/M_\odot)$ versus $\log_{10}(M_{\rm CSM}/M_\odot)$. Red circles, blue squares, and gray diamonds denote SNe Ibn, SNe Icn, and FBOTs, respectively. Error bars indicate the posterior 16th--84th percentile ranges transformed into the plotted physical units. The corresponding KDE-based OVLs between FBOTs and the combined Ibn/Icn distribution are ${\rm OVL}=0.51^{+0.10}_{-0.14}$ in panel~(a) and ${\rm OVL}=0.53^{+0.11}_{-0.13}$ in panel~(b).}
\label{fig:param_space_continuity}
\end{figure*}

We examine whether the observational continuity discussed above is also reflected in the fitted parameters of the unified \texttt{TransFit-CSM} model. Figure~\ref{fig:param_space_continuity} shows the posterior median parameters for the 25 retained events in two directly fitted parameter planes: the $R_{\rm CSM}$--$M_{\rm CSM}$ plane and the $M_{\rm ej}$--$M_{\rm CSM}$ plane. These projections were chosen because they describe the radial extent and mass scale of the optically thick CSM, as well as the relative amount of ejecta material participating in the interaction-powered emission.

Figure~\ref{fig:param_space_continuity}(a) shows that SNe Ibn, SNe Icn, and FBOTs occupy overlapping regions in the $R_{\rm CSM}$--$M_{\rm CSM}$ plane. FBOTs do not form a separate group in either CSM mass or CSM radius; instead, they are embedded within the broader distribution spanned by the full sample. This suggests that, within the adopted model, the optical light curves of FBOTs can be reproduced with CSM masses and radial scales comparable to those inferred for SNe Ibn/Icn.

A similar behavior is seen in the $M_{\rm ej}$--$M_{\rm CSM}$ plane shown in Figure~\ref{fig:param_space_continuity}(b). The three classes again show substantial overlap, indicating that the fitted ejecta and CSM masses do not define cleanly separated regions for SNe Ibn, SNe Icn, and FBOTs. The spread in this plane also reflects the fact that the light-curve timescale and luminosity are not controlled by a single mass parameter, but by a combination of ejecta mass, CSM mass, CSM extent, opacity, and heating efficiency.

Taken together, the two parameter-space projections show that the empirical similarity among the three classes is preserved after the light curves are mapped into the unified model. Under the common \texttt{TransFit-CSM} parameterization, SNe Ibn, SNe Icn, and FBOTs occupy overlapping regions of the fitted CSM and ejecta parameter space. This result supports the use of a dense CSM-interaction framework for comparing their optical light curves, while not requiring all events to share an identical progenitor channel or power-source composition.

\begin{figure}[t]
\centering
\includegraphics[width=\columnwidth]{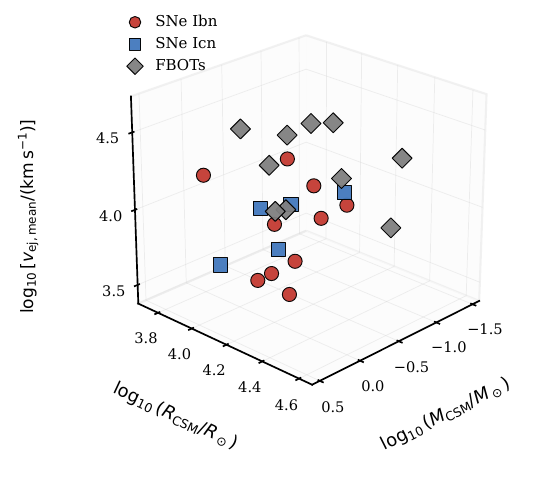}
\caption{Posterior median fitted parameters of the 25 retained events in the $M_{\rm CSM}$--$R_{\rm CSM}$--$v_{\rm ej,mean}$ space. All three axes are shown in logarithmic units. The velocity dimension highlights that FBOTs preferentially extend toward higher characteristic ejecta velocities. Symbols follow Figure~\ref{fig:param_space_continuity}.}
\label{fig:param_space_velocity_3d}
\end{figure}

As a complementary view of the fitted parameter space, Figure~\ref{fig:param_space_velocity_3d} includes the characteristic ejecta velocity,
\begin{equation}
v_{\rm ej,mean}=\left(\frac{2E_{\rm SN}}{M_{\rm ej}}\right)^{1/2}.
\end{equation}
This derived quantity combines the fitted explosion energy and ejecta mass and traces the kinetic-energy scale per unit ejecta mass. In this projection, FBOTs remain broadly embedded in the CSM mass--radius distribution, but they tend to extend toward higher $v_{\rm ej,mean}$. This suggests that the fast optical evolution of FBOTs is not associated with a distinct CSM mass or radius scale alone, but is also linked to the velocity or specific-energy scale of the ejecta.

\subsection{Implications for progenitor systems}
\label{sec:progenitors}

The fitted parameters also provide a way to discuss the possible
progenitor systems of SNe Ibn, SNe Icn, and FBOTs, although such an
interpretation should be made with caution. The light-curve fits mainly
constrain the mass and radial scale of the optically thick material that
participates in the thermal optical emission. They do not directly
determine the CSM composition, the binary status of the progenitor, or
the nature of any compact remnant. Nevertheless, the fitted
$R_{\rm CSM}$--$M_{\rm CSM}$ and $M_{\rm ej}$--$M_{\rm CSM}$ overlap
suggests that the three classes can share similar near-progenitor CSM
conditions, even if their spectroscopic classifications point to
different degrees of envelope stripping.

For the adopted posterior medians, the characteristic CSM radii are of
order $R_{\rm CSM}\sim 10^{15}\ {\rm cm}$. If this material was expelled
before core collapse with a characteristic velocity $v_{\rm w}$, the
corresponding look-back time is
\begin{equation}
t_{\rm pre}\simeq 0.22
\left(\frac{R_{\rm CSM}}{10^4\,R_\odot}\right)
\left(\frac{v_{\rm w}}{1000\ {\rm km\ s^{-1}}}\right)^{-1}
{\rm yr}.
\end{equation}
This estimate becomes about an order of magnitude longer for a slower
outflow with $v_{\rm w}\sim 100\ {\rm km\ s^{-1}}$. Therefore, if the
fitted CSM is interpreted as pre-explosion ejecta, the inferred
$M_{\rm CSM}\sim 10^{-1}$--$1\,M_\odot$ and compact radii favor an
episode of enhanced mass loss during the final years to decades before
explosion, rather than a long-lived, smooth, low-density wind.

For SNe Ibn, this interpretation is naturally consistent with explosions
of stripped helium-star progenitors embedded in He-rich, H-poor CSM
\citep[e.g.,][]{Hosseinzadeh2017,Dessart2022}. In our fits, many SNe Ibn
require CSM masses of order a few $0.1\,M_\odot$ to several $M_\odot$
within radii of order $10^{15}\ {\rm cm}$. Such compact and massive CSM
is more suggestive of eruptive or binary-mediated mass loss than of a
steady Wolf--Rayet wind alone. The relatively low fitted ejecta masses
for several objects also favor compact stripped progenitors, although
the broad posterior ranges prevent a unique distinction between a
single WR-like progenitor and a binary-stripped helium star.

For SNe Icn, the fitted CSM masses and radii are broadly comparable to
those of SNe Ibn, but their spectra require a more strongly stripped
environment dominated by C/O-rich and H/He-poor material
\citep[e.g.,][]{GalYam2022,Davis2023,Nagao2023}. This suggests that the
main difference between the Ibn and Icn progenitor systems may not be
the total amount or radial scale of dense CSM, but the chemical depth
from which the pre-supernova material was removed. In this sense, SNe
Icn may represent a more extreme stripping channel, possibly involving
compact CO/WO-like progenitors or binary systems capable of removing
most of the He-rich layers before explosion. The present optical fits do
not require SNe Icn to occupy a separate CSM mass scale from SNe Ibn;
instead, they indicate that similar interaction conditions can occur in
chemically different H-poor environments.

For FBOTs, the fitted parameters support a more heterogeneous
interpretation. Their posterior medians overlap with those of SNe Ibn
and SNe Icn in the CSM/ejecta parameter planes, indicating that at least
a subset of FBOT optical light curves can be produced by interaction
between fast ejecta and compact dense CSM. Compared with the Ibn/Icn
sample, however, FBOTs preferentially occupy the faster, more luminous,
and bluer end of the empirical distribution. Within the present
framework, this behavior can be produced by shorter effective diffusion
times, more compact emitting regions, higher shock-heating efficiencies,
or an additional inner heating component. The large effective $M_{\rm Ni}$
values inferred for several FBOTs should therefore not be interpreted
literally as radioactive nickel masses. Rather, they likely absorb
unmodeled inner power, such as central-engine activity, fallback
accretion, or other non-thermal components, especially for AT2018cow-like
events with luminous X-ray or radio emission \citep[e.g.,][]{Margutti2019,
Perley2019,Perley2021,Yao2022}.

Taken together, the fitted light curves favor a progenitor picture in
which SNe Ibn, SNe Icn, and at least some FBOTs arise from compact,
stripped systems surrounded by dense nearby CSM. The sequence from Ibn
to Icn can be understood primarily as a change in the chemical
composition and stripping depth of the CSM, while FBOTs extend the same
interaction parameter space toward shorter diffusion times and larger
specific energy input. This does not imply a single progenitor channel
for all FBOTs. Instead, it suggests that dense CSM interaction can
account for their thermal optical emission in a subset of cases, whereas
the most extreme FBOTs probably require an additional engine or
non-thermal component beyond the optical CSM-interaction model.

\section{Discussion and Conclusions}
\label{sec:discussion}

The empirical measurements and unified light-curve fits presented above
lead to a consistent but deliberately limited conclusion. SNe Ibn, SNe
Icn, and FBOTs are not fully mixed in all observed projections, but they
are also not separated into three isolated optical-transient populations.
The same-band light-curve planes show only limited overlap between
FBOTs and the combined Ibn/Icn sample, with KDE-based overlap
coefficients of $0.10^{+0.11}_{-0.07}$ in the
$\log_{10}(\nu L_{\nu,\rm peak,g})$--$\tau_{\rm rise,g}$ plane and
$0.10^{+0.10}_{-0.06}$ in the
$\Delta m_{10,g}$--$\tau_{\rm rise,g}$ plane. In contrast, the
peak-color distribution and the fitted CSM/ejecta parameter planes show
larger overlaps, with OVLs of $0.69^{+0.12}_{-0.17}$,
$0.51^{+0.10}_{-0.14}$, and $0.53^{+0.11}_{-0.13}$ for the
$(g-r)_{r,\max}$, $R_{\rm CSM}$--$M_{\rm CSM}$, and
$M_{\rm ej}$--$M_{\rm CSM}$ comparisons, respectively. These values
should be interpreted as descriptive measures of distributional overlap,
rather than as formal hypothesis tests.

This two-level behavior is physically informative. The relatively small
overlap in the light-curve planes reflects the fact that FBOTs
preferentially occupy the faster, more luminous, and often bluer end of
the observed distribution. The larger overlap in color and fitted
parameter space indicates that their optical emission can nevertheless
be described with CSM masses, CSM radii, and ejecta masses comparable to
those inferred for SNe Ibn/Icn. The velocity dimension adds a further
distinction: FBOTs tend to extend toward higher $v_{\rm ej,mean}$, even
though they do not occupy a separate CSM mass--radius region. Thus, the
difference between FBOTs and SNe Ibn/Icn in this framework is not simply
a difference in the total amount or radial scale of dense CSM. Instead,
FBOT-like optical light curves favor compact dense CSM together with a
larger ejecta velocity or specific-energy scale, with additional inner
power likely contributing in the most extreme events.

The fitted progenitor implications should therefore be stated in terms
of nearby dense CSM rather than a single progenitor channel. For SNe Ibn,
the inferred compact and massive CSM is naturally consistent with
stripped helium-star progenitors embedded in He-rich, H-poor material.
For SNe Icn, the similar fitted CSM mass and radius scale, together with
their C/O-rich and H/He-poor spectra, suggests a more deeply stripped
environment rather than a clearly distinct CSM mass scale. In this
sense, the Ibn--Icn comparison may trace differences in the chemical
depth of pre-supernova mass loss. FBOTs extend this interaction
parameter space toward shorter timescales and larger specific energy
input, but they should not be treated as a homogeneous progenitor class.
At least a subset of FBOTs can be compared with SNe Ibn/Icn through
their thermal optical emission, whereas the most extreme AT2018cow-like
events likely require additional central-engine, fallback, relativistic
outflow, or other non-thermal components.

Several limitations affect this interpretation. First, the empirical
comparison uses the observed $g$ band as a common reference band, rather
than a strictly identical rest-frame wavelength. This choice preserves a
larger sample but leaves residual effects from redshift, spectral
evolution, and $K$-corrections. Second, the peak-color comparison is
limited by the small number of objects with well-constrained standard
$g$- and $r$-band coverage around maximum light. Third, the
TransFit-CSM fits are not unique. Degeneracies among
$M_{\rm CSM}$, $R_{\rm CSM}$, $M_{\rm ej}$, $\kappa$,
$\epsilon_{\rm sh}$, and the effective inner heating component can
affect the inferred parameters. In particular, the fitted $M_{\rm Ni}$
should be regarded as an effective inner heating term in this unified
model, not as direct evidence that all events synthesize large
radioactive nickel masses. Finally, the present work models the thermal
optical emission only. Radio, X-ray, late-time spectroscopy, line
profiles, host environments, and pre-explosion mass-loss constraints are
needed to determine whether the optical overlap reflects a true physical
continuity among progenitor systems.

We conclude that a dense CSM-interaction framework provides a useful
common language for comparing the optical light curves of SNe Ibn, SNe
Icn, and at least a subset of FBOTs. The observed light-curve
differences are real: FBOTs preferentially occupy the rapidly evolving
and high-luminosity end of the distribution. However, their blue colors
and fitted CSM/ejecta parameters overlap substantially with those of
SNe Ibn/Icn. This supports a picture in which compact dense CSM
interaction contributes significantly to the thermal optical emission
across these classes, while chemical composition, stripping depth,
diffusion time, and possible additional inner power determine where each
event lies within the broader fast-blue-transient parameter space.

\begin{acknowledgements}
We thank Prof. Xiao-Feng Wang  for helpful discussions that improved this work.
This work was supported by the National Natural Science Foundation of China
(grant Nos. 12303047 and 12393811), the Natural Science Foundation of Hubei Province
(grant No. 2023AFB321), and the National Key R\&D Program of China
(grant No. 2021YFA0718500).
\end{acknowledgements}
\clearpage

\appendix

\section{Full-sample Multiband Light-curve Fits}
\label{app:full-sample-fits}

The multiband light-curve fits for all 25 retained events are presented in Figure~\ref{fig:full-sample-fits}.

\begin{figure*}[!ht]
    \figurenum{A1}
    \centering
    \includegraphics[
        width=0.98\textwidth,
        height=0.72\textheight,
        keepaspectratio
    ]{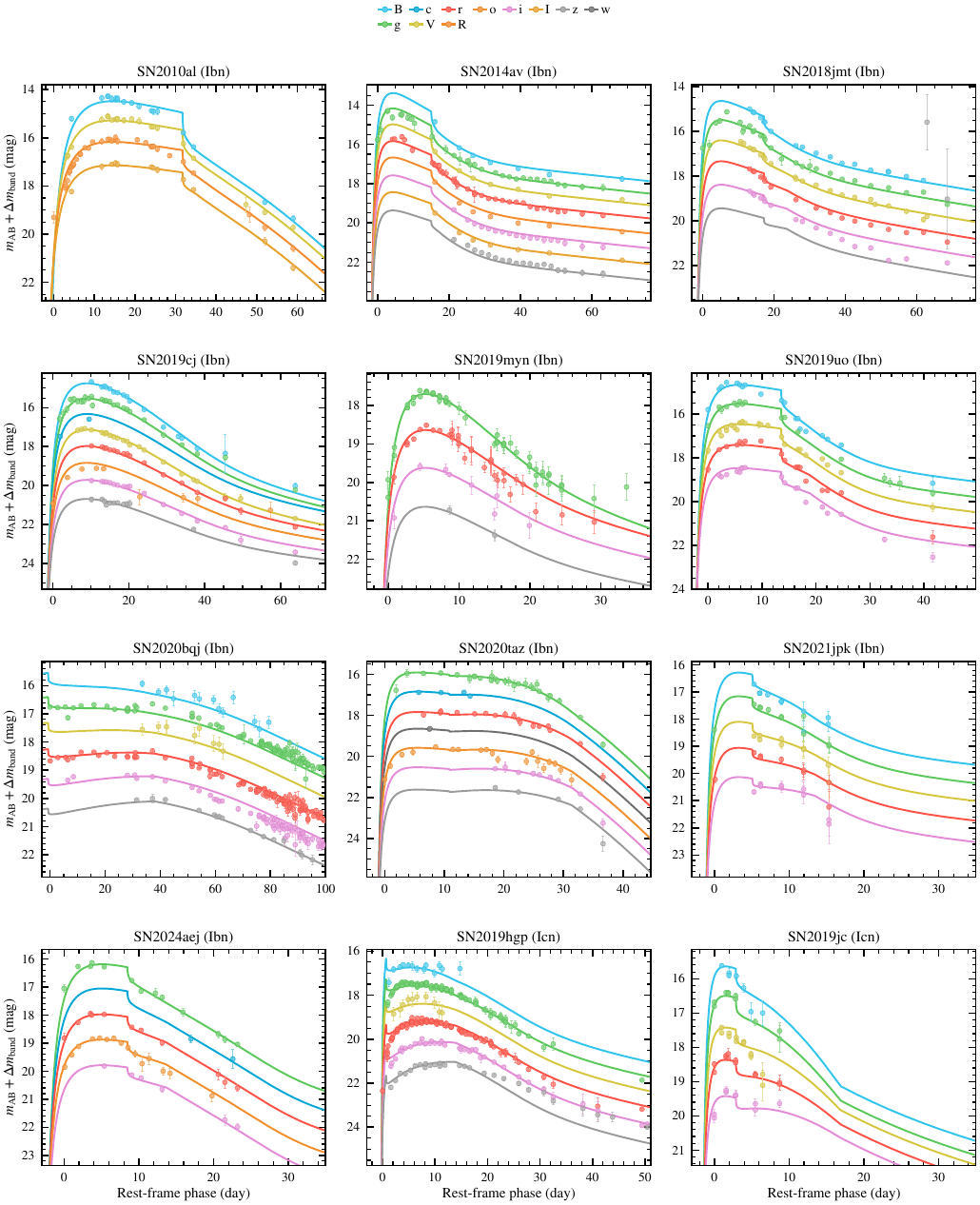}
    \caption{Multi-band observations and fits for the retained sample of SNe Ibn, SNe Icn, and FBOTs using the unified CSM+$^{56}$Ni model. The points represent the observed photometric data, and the solid curves show the best-fit model light curves in the corresponding bands. Colors indicate different photometric bands, as marked at the top of the figure. The ordinate is $m_{\rm AB}+\Delta m_{\rm band}$, where $\Delta m_{\rm band}$ is a fixed visual offset applied separately to each photometric band.}
    \label{fig:full-sample-fits}
\end{figure*}

\clearpage

\begin{figure*}[!t]
    \figurenum{A1}
    \centering
    \includegraphics[
        width=0.98\textwidth,
        height=0.86\textheight,
        keepaspectratio
    ]{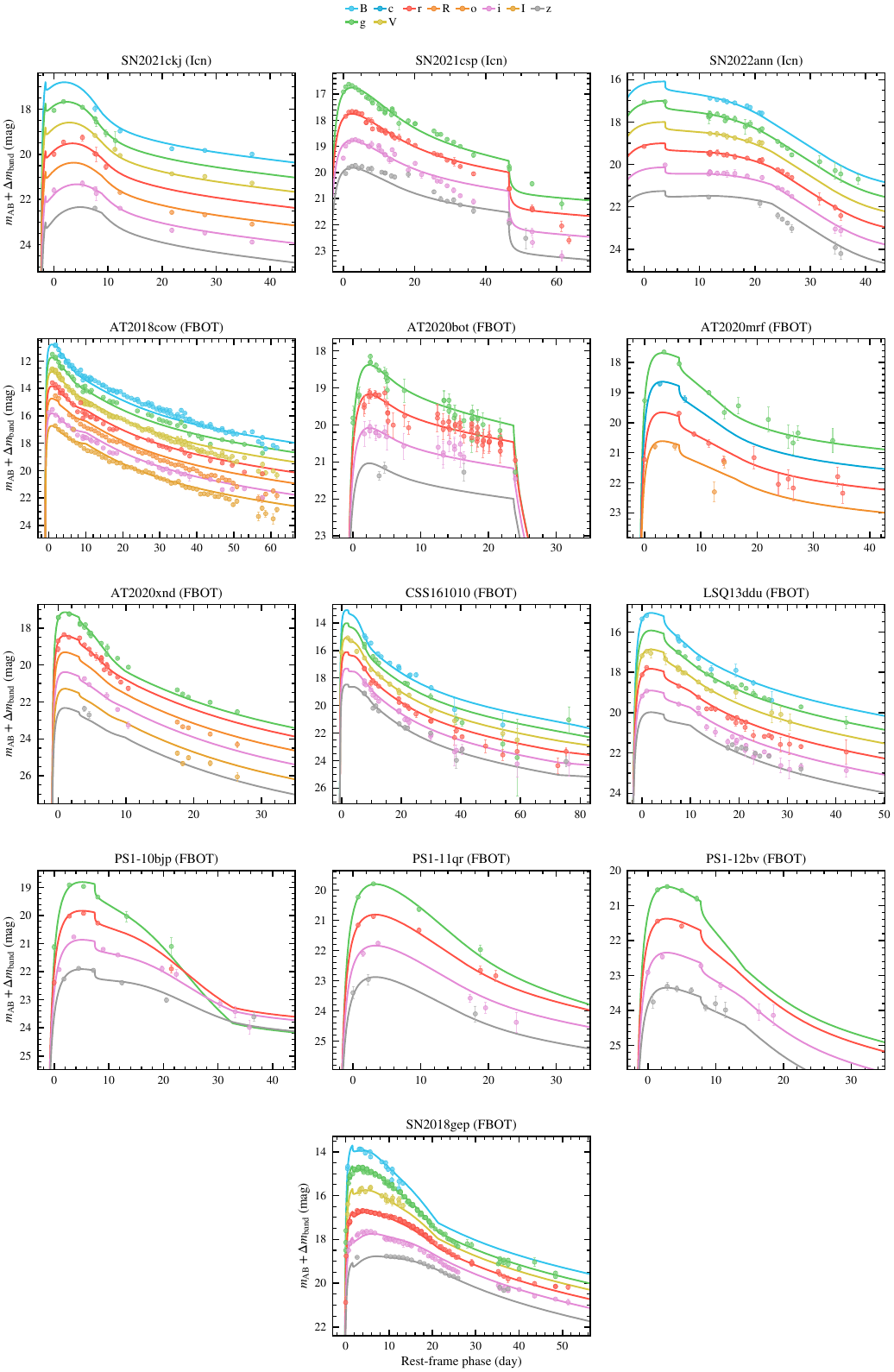}
    \caption{Continued.}
\end{figure*}

\clearpage

\section{KDE-based Overlap Measurements}
\label{app:kde-ovl}

We quantify the distributional overlap between FBOTs and the combined Ibn/Icn sample using Gaussian kernel density estimation (KDE; \citealt{Scott1992,Silverman1986}) and the overlapping coefficient \citep{Schmid2006}. For population $j$, containing $n_j$ objects with coordinates $\boldsymbol{x}_{j,i}$, the KDE is
\begin{equation}
\widehat{f}_j(\boldsymbol{x})
=
\frac{1}{n_j}
\sum_{i=1}^{n_j}
\mathcal{N}
\left(
\boldsymbol{x}\mid
\boldsymbol{x}_{j,i},
\boldsymbol{H}_j
\right),
\label{eq:kde-density}
\end{equation}
where $\boldsymbol{H}_j$ is the bandwidth matrix. The default bandwidth follows Scott's rule.

The overlap coefficient is defined as
\begin{equation}
{\rm OVL}
=
\int
\min
\left[
f_{\rm FBOT}(\boldsymbol{x}),
f_{\rm Ibn+Icn}(\boldsymbol{x})
\right]
\,d\boldsymbol{x}.
\label{eq:ovl-definition}
\end{equation}
The two densities are evaluated on the same grid and normalized before numerical integration. The OVL ranges from zero for non-overlapping densities to unity for identical densities.

Uncertainties in the OVL estimates are evaluated using 500 bootstrap realizations. In each realization, objects are resampled with replacement within each population. The empirical light-curve measurements and peak colors are perturbed according to their reported uncertainties, while the fitted parameter comparisons use joint draws from the source-level MCMC posteriors. We report the median and 16th--84th percentile range of the resulting OVL distribution. The relative ordering of the measured overlaps is unchanged when adopting Silverman's rule or bandwidths of 0.8 and 1.2 times the Scott value. The OVL is used as a descriptive measure of distributional overlap rather than as a formal hypothesis test.

\begin{figure*}[!ht]
    \figurenum{B1}
    \centering
    \includegraphics[
        width=0.98\textwidth,
        keepaspectratio
    ]{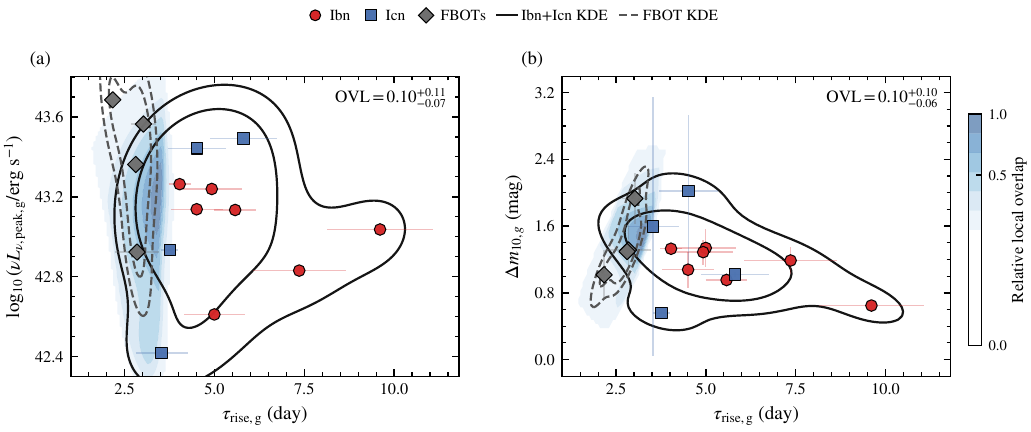}
    \caption{KDE representations of the same-band empirical light-curve planes shown in Figure~\ref{fig:gp_observables_continuity_2panel}. Panel~(a) shows $\log_{10}(\nu L_{\nu,\rm peak,g})$ versus $\tau_{\rm rise,g}$, and panel~(b) shows $\Delta m_{10,g}$ versus $\tau_{\rm rise,g}$. The solid black and dashed gray contours represent the combined Ibn/Icn and FBOT KDEs, respectively. The blue shading shows the local common density, $\min(f_{\rm Ibn+Icn},f_{\rm FBOT})$, normalized to its maximum for display. Symbols and error bars follow Figure~\ref{fig:gp_observables_continuity_2panel}. The contours and shading are calculated from the central measurements, whereas the annotated OVL values give the medians and 16th--84th percentile ranges obtained from the resampling analysis.}
    \label{fig:kde-overlap-empirical}
\end{figure*}

Figure~\ref{fig:kde-overlap-empirical} shows the KDE representations of the same-band empirical planes considered in Figure~\ref{fig:gp_observables_continuity_2panel}.

\clearpage

For the one-dimensional peak-color comparison in Figure~\ref{fig:gr_rmax_distribution}, Equation~(\ref{eq:ovl-definition}) reduces to
\begin{equation}
{\rm OVL}_{g-r}
=
\int
\min
\left[
f_{\rm FBOT}(c),
f_{\rm Ibn+Icn}(c)
\right]
\,dc,
\qquad
c=(g-r)_{r,\max}.
\label{eq:ovl-color}
\end{equation}
This calculation gives ${\rm OVL}_{g-r}=0.69^{+0.12}_{-0.17}$ for the peak-color subsample.

\begin{figure*}[!ht]
    \figurenum{B2}
    \centering
    \includegraphics[
        width=0.98\textwidth,
        keepaspectratio
    ]{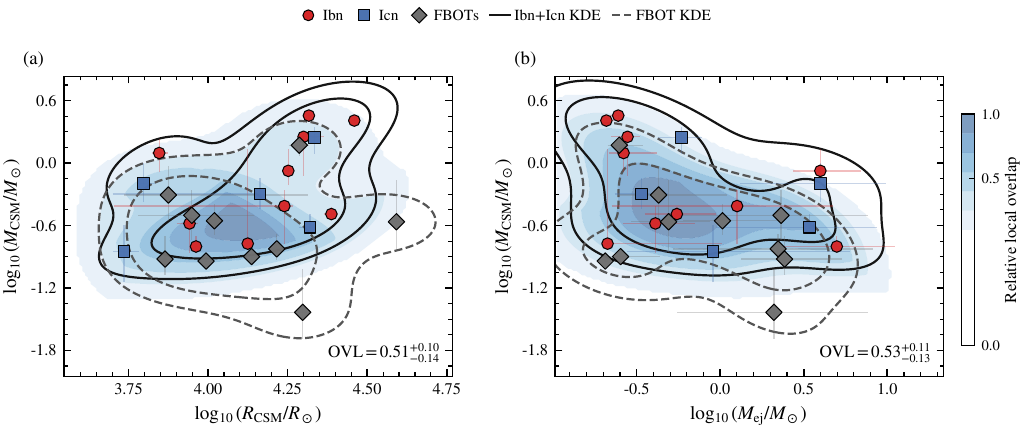}
    \caption{KDE representations of the fitted parameter planes shown in Figure~\ref{fig:param_space_continuity}. Panel~(a) shows $\log_{10}(R_{\rm CSM}/R_\odot)$ versus $\log_{10}(M_{\rm CSM}/M_\odot)$, and panel~(b) shows $\log_{10}(M_{\rm ej}/M_\odot)$ versus $\log_{10}(M_{\rm CSM}/M_\odot)$. The solid black and dashed gray contours represent the combined Ibn/Icn and FBOT KDEs, respectively. The blue shading shows the local common density, $\min(f_{\rm Ibn+Icn},f_{\rm FBOT})$, normalized to its maximum for display. Symbols and error bars follow Figure~\ref{fig:param_space_continuity}. The contours and shading are calculated from the central measurements, whereas the annotated OVL values give the medians and 16th--84th percentile ranges obtained from the resampling analysis.}
    \label{fig:kde-overlap-parameters}
\end{figure*}

Figure~\ref{fig:kde-overlap-parameters} shows the corresponding KDE representations of the fitted parameter planes in Figure~\ref{fig:param_space_continuity}. The KDEs and overlap integrals are evaluated in the logarithmic coordinates displayed in the figure.

\clearpage

\bibliography{sample701}{}

\begin{thebibliography}{}
\expandafter\ifx\csname natexlab\endcsname\relax\def\natexlab#1{#1}\fi
\providecommand{\url}[1]{\href{#1}{#1}}
\providecommand{\dodoi}[1]{doi:~\href{http://doi.org/#1}{\nolinkurl{#1}}}
\providecommand{\doeprint}[1]{\href{http://ascl.net/#1}{\nolinkurl{http://ascl.net/#1}}}
\providecommand{\doarXiv}[1]{\href{https://arxiv.org/abs/#1}{\nolinkurl{https://arxiv.org/abs/#1}}}

% type= article
\bibitem[{S. {Ambikasaran} {et~al.}(2016){Ambikasaran}, {Foreman-Mackey}, {Greengard}, {Hogg}, \& {O'Neil}}]{Ambikasaran2016}
{Ambikasaran}, S., {Foreman-Mackey}, D., {Greengard}, L., {Hogg}, D.~W., \& {O'Neil}, M. 2016, \bibinfo{title}{{Fast Direct Methods for Gaussian Processes},} IEEE Transactions on Pattern Analysis and Machine Intelligence, 38, 252, \dodoi{10.1109/TPAMI.2015.2448083}

% type= article
\bibitem[{E. {Chatzopoulos} {et~al.}(2012){Chatzopoulos}, {Wheeler}, \& {Vink{\'o}}}]{Chatzopoulos2012}
{Chatzopoulos}, E., {Wheeler}, J.~C., \& {Vink{\'o}}, J. 2012, \bibinfo{title}{{Generalized Semi-analytical Models of Supernova Light Curves},} \apj, 746, 121, \dodoi{10.1088/0004-637X/746/2/121}

% type= article
\bibitem[{E. {Chatzopoulos} {et~al.}(2013){Chatzopoulos}, {Wheeler}, {Vink{\'o}}, {Horv{\'a}th}, \& {Nagy}}]{Chatzopoulos2013}
{Chatzopoulos}, E., {Wheeler}, J.~C., {Vink{\'o}}, J., {Horv{\'a}th}, Z.~L., \& {Nagy}, A. 2013, \bibinfo{title}{{Analytical Light Curve Models of Superluminous Supernovae: {$\chi^2$}-Minimization of Parameter Fits},} \apj, 773, 76, \dodoi{10.1088/0004-637X/773/1/76}

% type= article
\bibitem[{P. {Clark} {et~al.}(2020){Clark} {et~al.}}]{Clark2020}
{Clark}, P., {et~al.} 2020, \bibinfo{title}{{LSQ13ddu: A Rapidly Evolving Stripped-envelope Supernova with Early Circumstellar Interaction Signatures},} \mnras, 492, 2208, \dodoi{10.1093/mnras/stz3598}

% type= article
\bibitem[{K.~W. {Davis} {et~al.}(2023){Davis} {et~al.}}]{Davis2023}
{Davis}, K.~W., {et~al.} 2023, \bibinfo{title}{{SN 2022ann: A Type Icn Supernova from a Dwarf Galaxy That Reveals Helium in Its Circumstellar Environment},} \mnras, 523, 2530, \dodoi{10.1093/mnras/stad1433}

% type= article
\bibitem[{L. {Dessart} {et~al.}(2022){Dessart}, {Hillier}, \& {Kuncarayakti}}]{Dessart2022}
{Dessart}, L., {Hillier}, D.~J., \& {Kuncarayakti}, H. 2022, \bibinfo{title}{{Helium stars exploding in circumstellar material and the origin of Type Ibn supernovae},} \aap, 658, A130, \dodoi{10.1051/0004-6361/202142436}

% type= article
\bibitem[{M.~R. {Drout} {et~al.}(2014){Drout} {et~al.}}]{Drout2014}
{Drout}, M.~R., {et~al.} 2014, \bibinfo{title}{{Rapidly Evolving and Luminous Transients from Pan-STARRS1},} \apj, 794, 23, \dodoi{10.1088/0004-637X/794/1/23}

% type= article
\bibitem[{A. {Gal-Yam} {et~al.}(2022){Gal-Yam} {et~al.}}]{GalYam2022}
{Gal-Yam}, A., {et~al.} 2022, \bibinfo{title}{{A WC/WO Star Exploding within an Expanding Carbon--Oxygen--Neon Nebula},} \nat, 601, 201, \dodoi{10.1038/s41586-021-04155-1}

% type= article
\bibitem[{A. {Gangopadhyay} {et~al.}(2020){Gangopadhyay} {et~al.}}]{Gangopadhyay2020}
{Gangopadhyay}, A., {et~al.} 2020, \bibinfo{title}{{Flash Ionization Signatures in the Type Ibn Supernova SN 2019uo},} \apj, 889, 170, \dodoi{10.3847/1538-4357/ab6328}

% type= article
\bibitem[{C.~P. {Guti{\'e}rrez} {et~al.}(2024){Guti{\'e}rrez} {et~al.}}]{Gutierrez2024}
{Guti{\'e}rrez}, C.~P., {et~al.} 2024, \bibinfo{title}{{CSS 161010: A Luminous Fast Blue Optical Transient with Broad Blueshifted Hydrogen Lines},} \apj, 977, 162, \dodoi{10.3847/1538-4357/ad89a5}

% type= article
\bibitem[{A.~Y.~Q. {Ho} {et~al.}(2019){Ho} {et~al.}}]{Ho2019}
{Ho}, A. Y.~Q., {et~al.} 2019, \bibinfo{title}{{Evidence for Late-stage Eruptive Mass Loss in the Progenitor to SN2018gep, a Broad-lined Ic Supernova: Pre-explosion Emission and a Rapidly Rising Luminous Transient},} \apj, 887, 169, \dodoi{10.3847/1538-4357/ab55ec}

% type= article
\bibitem[{A.~Y.~Q. {Ho} {et~al.}(2023){Ho} {et~al.}}]{Ho2023}
{Ho}, A. Y.~Q., {et~al.} 2023, \bibinfo{title}{{A Search for Extragalactic Fast Blue Optical Transients in ZTF and the Rate of AT2018cow-like Transients},} \apj, 949, 120, \dodoi{10.3847/1538-4357/acc533}

% type= article
\bibitem[{G. {Hosseinzadeh} {et~al.}(2017){Hosseinzadeh} {et~al.}}]{Hosseinzadeh2017}
{Hosseinzadeh}, G., {et~al.} 2017, \bibinfo{title}{{Type Ibn Supernovae Show Photometric Homogeneity and Spectral Diversity at Maximum Light},} \apj, 836, 158, \dodoi{10.3847/1538-4357/836/2/158}

% type= article
\bibitem[{C. {Inserra}(2019){Inserra}}]{Inserra2019}
{Inserra}, C. 2019, \bibinfo{title}{{Observational properties of extreme supernovae},} Nature Astronomy, 3, 697, \dodoi{10.1038/s41550-019-0854-4}

% type= article
\bibitem[{J.-a. {Jiang} {et~al.}(2022){Jiang}, {Yasuda}, {Maeda}, {Tominaga}, {Doi}, {Ivezi{\'c}}, {Yoachim}, {Uno}, {Moriya}, {Kumar}, {Pan}, {Tanaka}, {Tanaka}, {Nomoto}, {Jha}, {Ruiz-Lapuente}, {Jones}, {Shigeyama}, {Suzuki}, {Kokubo}, {Furusawa}, {Miyazaki}, {Connolly}, {Sahu}, \& {Anupama}}]{Jiang2022}
{Jiang}, J.-a., {Yasuda}, N., {Maeda}, K., {et~al.} 2022, \bibinfo{title}{{MUSSES2020J: The Earliest Discovery of a Fast Blue Ultraluminous Transient at Redshift 1.063},} \apjl, 933, L36, \dodoi{10.3847/2041-8213/ac7390}

% type= article
\bibitem[{H. {Jin} {et~al.}(2026){Jin} {et~al.}}]{Jin2026}
{Jin}, H., {et~al.} 2026, \bibinfo{title}{{Type Ib Supernovae are bluer than Type Ic Supernovae},} arXiv e-prints, arXiv:2605.01200, \dodoi{10.48550/arXiv.2605.01200}

% type= article
\bibitem[{E.~C. {Kool} {et~al.}(2021){Kool} {et~al.}}]{Kool2021}
{Kool}, E.~C., {et~al.} 2021, \bibinfo{title}{{SN 2020bqj: A Type Ibn Supernova with a Long-lasting Peak Plateau},} \aap, 652, A136, \dodoi{10.1051/0004-6361/202039137}

% type= article
\bibitem[{J.-F. {Liu} {et~al.}(2022){Liu}, {Zhu}, {Liu}, {Yu}, \& {Zhang}}]{Liu2022}
{Liu}, J.-F., {Zhu}, J.-P., {Liu}, L.-D., {Yu}, Y.-W., \& {Zhang}, B. 2022, \bibinfo{title}{{Magnetar Engines in Fast Blue Optical Transients and Their Connections with SLSNe, SNe Ic-BL, and lGRBs},} \apjl, 935, L34, \dodoi{10.3847/2041-8213/ac86d2}

% type= article
\bibitem[{L.-D. {Liu} {et~al.}(2025){Liu}, {Zhang}, {Yu}, {Du}, {Li}, {Wu}, \& {Dai}}]{Liu2025TransFit}
{Liu}, L.-D., {Zhang}, Y.-H., {Yu}, Y.-W., {et~al.} 2025, \bibinfo{title}{{TransFit: An Efficient Framework for Transient Light-curve Fitting with Time-dependent Radiative Diffusion},} \apj, 992, 20, \dodoi{10.3847/1538-4357/adfed6}

% type= article
\bibitem[{K. {Maeda} \& T.~J. {Moriya}(2022){Maeda} \& {Moriya}}]{Maeda2022}
{Maeda}, K., \& {Moriya}, T.~J. 2022, \bibinfo{title}{{Properties of Type Ibn Supernovae: Implications for the Progenitor Evolution and the Origin of a Population of Rapid Transients},} \apj, 927, 25, \dodoi{10.3847/1538-4357/ac4672}

% type= article
\bibitem[{R. {Margutti} {et~al.}(2019){Margutti} {et~al.}}]{Margutti2019}
{Margutti}, R., {et~al.} 2019, \bibinfo{title}{{An Embedded X-Ray Source Shines through the Aspherical AT2018cow: Revealing the Inner Workings of the Most Luminous Fast-evolving Optical Transients},} \apj, 872, 18, \dodoi{10.3847/1538-4357/aafa01}

% type= article
\bibitem[{T. {Nagao} {et~al.}(2023){Nagao} {et~al.}}]{Nagao2023}
{Nagao}, T., {et~al.} 2023, \bibinfo{title}{{Photometry and Spectroscopy of the Type Icn Supernova 2021ckj: The Diverse Properties of the Ejecta and Circumstellar Matter of Type Icn Supernovae},} \aap, 673, A27, \dodoi{10.1051/0004-6361/202346084}

% type= article
\bibitem[{M. {Nicholl} {et~al.}(2015){Nicholl} {et~al.}}]{Nicholl2015}
{Nicholl}, M., {et~al.} 2015, \bibinfo{title}{{On the Diversity of Superluminous Supernovae: Ejected Mass as the Dominant Factor},} \mnras, 452, 3869, \dodoi{10.1093/mnras/stv1522}

% type= article
\bibitem[{A. {Pastorello} {et~al.}(2015){Pastorello} {et~al.}}]{Pastorello2015a}
{Pastorello}, A., {et~al.} 2015, \bibinfo{title}{{Massive Stars Exploding in a He-rich Circumstellar Medium -- IV. Transitional Type Ibn Supernovae},} \mnras, 449, 1921, \dodoi{10.1093/mnras/stu2745}

% type= article
\bibitem[{A. {Pastorello} {et~al.}(2016){Pastorello} {et~al.}}]{Pastorello2016}
{Pastorello}, A., {et~al.} 2016, \bibinfo{title}{{Massive Stars Exploding in a He-rich Circumstellar Medium -- IX. SN 2014av, and Characterization of Type Ibn SNe},} \mnras, 456, 853, \dodoi{10.1093/mnras/stv2634}

% type= article
\bibitem[{C. {Pellegrino} {et~al.}(2022{\natexlab{a}}){Pellegrino} {et~al.}}]{Pellegrino2022a}
{Pellegrino}, C., {et~al.} 2022{\natexlab{a}}, \bibinfo{title}{{Circumstellar Interaction Powers the Light Curves of Luminous Rapidly Evolving Optical Transients},} \apj, 926, 125, \dodoi{10.3847/1538-4357/ac3e63}

% type= article
\bibitem[{C. {Pellegrino} {et~al.}(2022{\natexlab{b}}){Pellegrino} {et~al.}}]{Pellegrino2022b}
{Pellegrino}, C., {et~al.} 2022{\natexlab{b}}, \bibinfo{title}{{The Diverse Properties of Type Icn Supernovae Point to Multiple Progenitor Channels},} \apj, 938, 73, \dodoi{10.3847/1538-4357/ac8ff6}

% type= article
\bibitem[{D.~A. {Perley} {et~al.}(2019){Perley} {et~al.}}]{Perley2019}
{Perley}, D.~A., {et~al.} 2019, \bibinfo{title}{{The Fast, Luminous Ultraviolet Transient AT2018cow: Extreme Supernova, or Disruption of a Star by an Intermediate-mass Black Hole?},} \mnras, 484, 1031, \dodoi{10.1093/mnras/sty3420}

% type= article
\bibitem[{D.~A. {Perley} {et~al.}(2021){Perley} {et~al.}}]{Perley2021}
{Perley}, D.~A., {et~al.} 2021, \bibinfo{title}{{Real-time Discovery of AT2020xnd: A Fast, Luminous Ultraviolet Transient with Minimal Radioactive Ejecta},} \mnras, 508, 5138, \dodoi{10.1093/mnras/stab2785}

% type= article
\bibitem[{D.~A. {Perley} {et~al.}(2022){Perley} {et~al.}}]{Perley2022}
{Perley}, D.~A., {et~al.} 2022, \bibinfo{title}{{The Type Icn SN 2021csp: Implications for the Origins of the Fastest Supernovae and the Fates of Wolf-Rayet Stars},} \apj, 927, 180, \dodoi{10.3847/1538-4357/ac478e}

% type= article
\bibitem[{M. {Pursiainen} {et~al.}(2018){Pursiainen} {et~al.}}]{Pursiainen2018}
{Pursiainen}, M., {et~al.} 2018, \bibinfo{title}{{Rapidly Evolving Transients in the Dark Energy Survey},} \mnras, 481, 894, \dodoi{10.1093/mnras/sty2309}

% type= article
\bibitem[{F. Schmid \& A. Schmidt(2006)Schmid \& Schmidt}]{Schmid2006}
Schmid, F., \& Schmidt, A. 2006, \bibinfo{title}{Nonparametric Estimation of the Coefficient of Overlapping: Theory and Empirical Application,} Computational Statistics \& Data Analysis, 50, 1583, \dodoi{10.1016/j.csda.2005.01.014}

% type= book
\bibitem[{D.~W. Scott(1992)Scott}]{Scott1992}
Scott, D.~W. 1992, Multivariate Density Estimation: Theory, Practice, and Visualization (New York: John Wiley \& Sons), \dodoi{10.1002/9780470316849}

% type= book
\bibitem[{B.~W. Silverman(1986)Silverman}]{Silverman1986}
Silverman, B.~W. 1986, Density Estimation for Statistics and Data Analysis (London: Chapman and Hall)

% type= article
\bibitem[{Z.-Y. {Wang} {et~al.}(2024){Wang} {et~al.}}]{Wang2024}
{Wang}, Z.-Y., {et~al.} 2024, \bibinfo{title}{{Massive Stars Exploding in a He-rich Circumstellar Medium -- X. Flash Spectral Features in the Type Ibn SN 2019cj and Observations of SN 2018jmt},} \aap, 691, A156, \dodoi{10.1051/0004-6361/202451131}

% type= article
\bibitem[{Z.-Y. {Wang} {et~al.}(2025){Wang} {et~al.}}]{Wang2025}
{Wang}, Z.-Y., {et~al.} 2025, \bibinfo{title}{{Massive Stars Exploding in a He-rich Circumstellar Medium -- XI. Diverse Evolution of Five Ibn SNe 2020nxt, 2020taz, 2021bbv, 2023utc, and 2024aej},} \aap, 700, A156, \dodoi{10.1051/0004-6361/202554768}

% type= article
\bibitem[{D. {Xiang} {et~al.}(2021){Xiang} {et~al.}}]{Xiang2021}
{Xiang}, D., {et~al.} 2021, \bibinfo{title}{{The Peculiar Transient AT2018cow: A Possible Origin of a Type Ibn/IIn Supernova},} \apj, 910, 42, \dodoi{10.3847/1538-4357/abdeba}

% type= article
\bibitem[{Y. {Yao} {et~al.}(2022){Yao} {et~al.}}]{Yao2022}
{Yao}, Y., {et~al.} 2022, \bibinfo{title}{{The X-Ray and Radio Loud Fast Blue Optical Transient AT2020mrf: Implications for an Emerging Class of Engine-driven Massive Star Explosions},} \apj, 934, 104, \dodoi{10.3847/1538-4357/ac7a41}

% type= article
\bibitem[{Y.-W. {Yu} {et~al.}(2015){Yu}, {Li}, \& {Dai}}]{Yu2015}
{Yu}, Y.-W., {Li}, S.-Z., \& {Dai}, Z.-G. 2015, \bibinfo{title}{{Rapidly Evolving and Luminous Transients Driven by Newly Born Neutron Stars},} \apjl, 806, L6, \dodoi{10.1088/2041-8205/806/1/L6}

% type= article
\bibitem[{Y.-H. {Zhang} {et~al.}(2026){Zhang}, {Liu}, {Du}, {Wu}, {Li}, \& {Yu}}]{Zhang2026TransFitCSM}
{Zhang}, Y.-H., {Liu}, L.-D., {Du}, Z.-X., {et~al.} 2026, \bibinfo{title}{{TransFit-CSM: A Fast, Physically Consistent Framework for Interaction-powered Transients},} \apj, 999, 186, \dodoi{10.3847/1538-4357/ae434a}

\end{thebibliography}
\bibliographystyle{aasjournalv7}
\end{document}